\documentclass[conference]{IEEEtran}
\IEEEoverridecommandlockouts
\usepackage{cite}
\usepackage{amsmath,amssymb,amsfonts}
\usepackage[compact]{titlesec}

\titlespacing{\section}{12pt}{*0}{*0}

\usepackage{algorithmic}
\usepackage{graphicx}
\usepackage{textcomp}
\usepackage{xcolor}
\def\BibTeX{{\rm B\kern-.05em{\sc i\kern-.025em b}\kern-.08em
    T\kern-.1667em\lower.7ex\hbox{E}\kern-.125emX}}

\begin{document}

\title{OMINACS: Online ML-Based IoT Network Attack Detection and Classification System\\ }

\author{\IEEEauthorblockN{1\textsuperscript{st} Diego Abreu}
\IEEEauthorblockA{
\textit{Federal University of Pará (UFPA)}\\
Belém, Brazil \\
diego.abreu@itec.ufpa.br}
\and
\IEEEauthorblockN{2\textsuperscript{nd} Antonio Abelém}
\IEEEauthorblockA{
\textit{Federal University of Pará (UFPA)}\\
Belém, Brazil\\
abelem@ufpa.br}
\thanks{The authors would like to thank the São Paulo Research Foundation (FAPESP) for the financial support through Grant 2020/04031-1. }

}

\maketitle

\begin{abstract}

Several Machine Learning (ML) methodologies have been proposed to improve security in Internet Of Things (IoT) networks and reduce the damage caused by the action of malicious agents. However, detecting and classifying attacks with high accuracy and precision is still a major challenge. This paper proposes an online attack detection and network traffic classification system, which combines stream Machine Learning, Deep Learning, and Ensemble Learning technique. Using multiple stages of data analysis, the system can detect the presence of malicious traffic flows and classify them according to the type of attack they represent. Furthermore, we show how to implement this system both in an IoT network and from an ML point of view. The system was evaluated in three IoT network security datasets, in which it obtained accuracy and precision above 90\% with a reduced false alarm rate.
\end{abstract}

\begin{IEEEkeywords}
Machine Learning, Network Security, Internet of Things.
\end{IEEEkeywords}

\section{Introduction}
\label{sec:intro}
With the rise of the Internet of Things (IoT), the number and diversity of threats to the security of computer networks are increasing\cite{introsurvey1}. Thus, detecting attacks and protecting the network has become a very challenging task for security mechanisms, such as the Intrusion Detection System (IDS) and Intrusion Prevention System (IPS) \cite{introsurvey2}. The main challenges faced in monitoring and preventing these attacks are the large amount of data generated by the IoT devices and the costs and time of analysis and processing involved. With a large number of IoT devices and with the heterogeneity of the network, attacks can come from multiple sources, generating a constant flow of data to be analyzed. \cite{lobato2016}. This results in a long delay in detecting attacks and in the number of false alarms generated by current monitoring systems \cite{introsurvey1}.

In this context, several Machine Learning (ML) techniques have been proposed to detect the presence of malicious agents in the network, which can mean the occurrence of an attack. In particular, Deep Learning techniques and techniques based on Ensemble Learning have obtained significant results in terms of precision and accuracy \cite{vsmlsurvey}.

However, many of these proposals tackle attack detection as an offline learning task, in which learning models are first trained and fine-tuned, only and then applied to a test detection system. These approaches tend to disregard the dynamic and adversarial behavior network attacks, in which the concept drift occurs, both in changes in the statistics of the learning and prediction target and in the network features. Thus, the models trained offline quickly become obsolete, which reduces the system accuracy. Furthermore, it is not only relevant to detect malicious activity on the network but it is also crucial to identify and classify the attacks. According to the specific type of attack identified, various actions can be taken to stop and prevent the future occurrence of this attack \cite{introsurvey2}.

Our proposal seeks to detect IoT attacks and classify them according to classes of attacks. Our approach consists of multiple online stages of network flow data analysis, applying specific ML techniques according to the most appropriate task for these stages. Our research hypothesis is that it is possible to develop a system capable of detecting and classifying dynamic attacks and combining the advantages of stream ML, Deep Learning and Ensemble Learning. The rest of the paper is organized as follows: section II presents the theoretical background; sections III and IV present the related works and the proposed system; section V presents the results of experiments and discussions; finally, section VI concludes the work and points out future works.

\section{Machine Learning Background}
This section presents the theoretical foundation for the system proposed in this work. The proposed system combines Ensemble ML, Deep Learning, and stream ML, seeking to obtain the main advantages of each methodology.

\textbf{Ensemble Machine Learning:}  is a methodology that uses different classification algorithms to create prediction models and combine individual results to obtain an optimized collective result \cite{major}. Among the different approaches of Ensemble, we can highlight the Majority Voting \cite{major}. In this approach, each model makes its own prediction about the association of each instance to a class, which represents a vote. The class that receives the highest number of votes is then chosen for the analyzed instance and the final prediction is determined according to the majority of votes cast. This process is finished when all instances in the dataset have been classified.

In the proposed system three supervised ML methods are used as the Ensemble classifiers: Random Forest, K-Nearest Neighbor (kNN), and Support Vector Machine (SVM). These classifiers were chosen because they are often used in the context of attack detection and also because they work differently from each other, which improves their use within the Majority Vote approach \cite{vsmlsurvey}. 

\textbf{Deep Learning:} in the context of network attack detection, one of the most used Deep Learning techniques is the Long Short-Term Memory Networks (LSTMs) \cite{LSTM}. While some DL can work with long sequences of data, such as time series, there is a limitation in retrieving information from previous sequences. LSTMs seek to mitigate this limitation of using long-term memory cells, with specific gates for information control. Thus LSTMs are suitable to be used to identify patterns in a normal traffic network and compare it with known attack classes.

\textbf{Stream Machine Learning:} unlike traditional offline Machine Learning, in which the whole dataset is available to the learning method at the time of training, in Stream ML the data is classified at the moment it arrives at the predictor. Thus, stream ML processes data one instance, or set of instances, at a time. Therefore, it is possible to use this approach to perform data classification in real-time, as the data arrives at the learning system.

During the processing of data in stream ML, it is common to see significant changes in the values of attributes, named concept drifts. In the context of an IoT network,  concept drifts can be caused either by changes in the normal behavior of the network or by the action of malicious agents in network attacks. For this reason, the learning model tends to lose part of its predictive capacity, making it necessary to re-train the model. The stream ML algorithms look for ways to be more robust to these concept changes, in order to avoid decreasing the accuracy and precision, thus avoiding model retraining. In this work we will use the Hoeffding Adaptive Tree (HAT) \cite{HAT}, which is a tree-based stream ML that adapts and learns from the changing data streams over time and has been proven to be very effective in detecting and classifying network attacks \cite{nethat}.

In Stream ML, the first data to arrive at the system is used to train the learning model. This model is then applied to the next data sequence, performing the prediction and classification of the data. When a concept drift is detected, it is necessary to perform a new training of the model, so that it can be adapted to the new state of the data flow. With the adapted model, the data prediction for the next data sequence is then performed again. This process continues until all data is read. The feature selection process is performed at each training or re-training of the model. The technique used has to be fast enough not to delay detection and should also help to create more robust models, able to adjust to changes in concept, and thus avoid retraining. In our system, the feature selection is approached in an integrated way both in the online detection of the models and in the classification by type of attacks, using the feature selection by clustering \cite{abreu2020}, an unsupervised technique that avoids \textit{overfitting} the majority classes by disregarding the data labels in the selection of the feature set.

\section{Related Works}
In Lucas et al. (2021) \cite{stack} a multi-stage attack detection system based on Ensemble Learning is proposed. The authors combine several classification methods to get the best configuration of the classifiers' hyperparameters. The system is then tested using the CICID17 \cite{cid17} dataset, obtaining results of accuracy and precision close to 100\%. However, despite these good results, the paper does not analyze the time used to generate these optimized settings, and all of the processing is done offline. In our proposal, a multi-stage attack detection and classification system is also used, however, unlike that proposed in Lucas et al (2021), the time and impact that the generation and application of models cause on the system will also be evaluated. In addition, we will also use the Ensemble approach as a part of our detection system.

In Tian et al. (2021) \cite{twostagesdn} a two-stage attack detection and classification approach is proposed for the context of Software Defined Networks (SDN). The authors use the first stage to select the network features using a bio-inspired method to obtain an optimized set of the most relevant features. In the second stage, with the reduced set of features, is applied an Ensemble classifier that combines a Decision Tree, an MLP Neural Network (Multilayer Perceptron) and the k-Nearest Neighbor. The proposed system is tested in the NSL-KDD \cite{NSL} and UNSW-NB15 \cite{Moustafa1} datasets, obtaining results close to 100\% for the binary detection, which classify the network traffic data between malicious and benign. However, in the multiclass classification, which considers the type of attack, the system only has good results for attacks with a high amount of data available for training. This is due both to the unbalance of the classes and the fact that the features set was optimized to detect the attack class itself but not specific attacks as the target. This caused the system to overfit the model towards the majority class. In our work, the feature selection is done continuously within the ML process, dynamically selecting the features according to the incoming data stream and thus not requiring a previous stage of feature analysis, which, in an online system, could cause a significant increase in attack detection time.


In our previous research \cite{abreu2022} an ML pipeline based on stream learning is presented, which integrates dynamic feature selection, and fast decision tree models, to detect and classify network attacks on stream ML. The ML pipeline was tested in the NLS-KDD, UNSW-NB15, and CICIDS17 datasets, showing high accuracy results, with low false alarms.
Now, with the focus on IoT networks, this ML pipeline combined with the fog, edge, and cloud network view, creating the OMINACS system. The idea is to apply the ML pipeline into a system that can be distributed through an IoT network, considering the multiple network layers such as the edge, fog, and cloud devices.

\section{OMINACS: Online ML-Based IoT Network Attack Detection and Classification System}

\subsection{The Machine Learning Pipeline}
Essentially the proposed system can be understood as a sequence of Machine Learning processes. The system takes as input the data of the network traffic and, as an output, it presents the data classified according to traffic classes and the system evaluation metrics. Figure 1 shows how the system works. The system has four stages that perform network traffic classification and attack detection.

\begin{figure*}[h]
  \centering
    \includegraphics[width=2\columnwidth,
    height= 12.0cm]{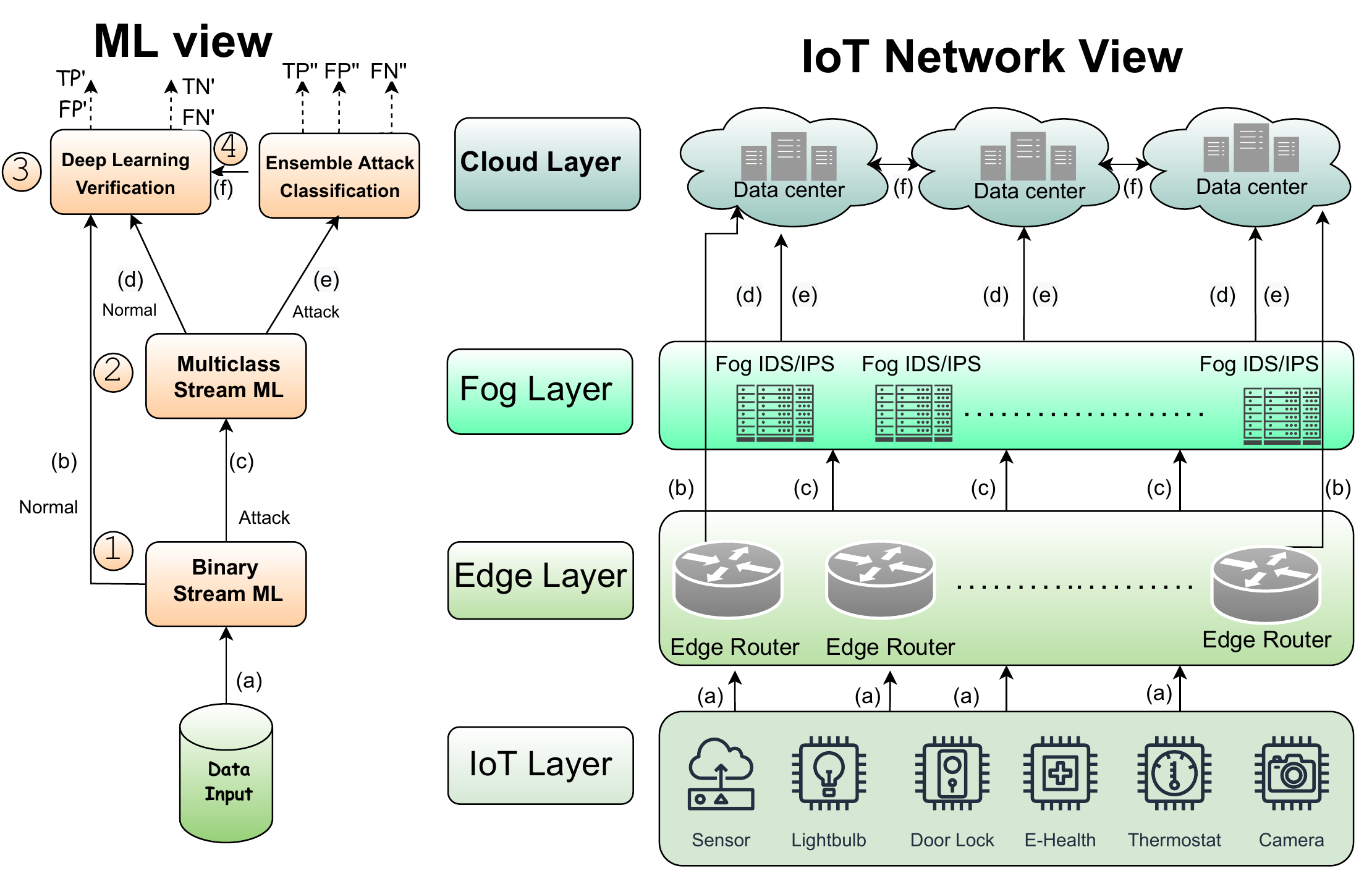}
   \label{propostasbrc22}
  \caption{OMINACS: Online ML-Based IoT Network Attack Detection and Classification System.}
\end{figure*}

Stage 1 is the first part of the system. The network data flow is analyzed and classified between traffic data considered as normal or attack data. Thus, for this stage, an ML-based detection system is implemented using the HAT stream ML method. At this stage, the HAT has a binary target and classifies the data into two classes: normal traffic and attack traffic. At each training or re-training of the model, the feature selection technique \cite{abreu2020} is applied, improving the performance of the HAT. Classified instances are streamed to stage 2 or 3, as they are labeled by the HAT. The data classified as an attack is sent to stage 2, and data classified as normal is sent to stage 3.

Similar to stage 1, in stage 2, the data is classified according to the continuous flow of incoming data from network traffic. In this step, stream ML is used to classify data that was already flagged in stage 1 as an attack, now between types of attacks. Thus, the multiclass version of HAT, which targets multiple data classes,  was implemented. At the end of this stage, the data is classified into different classes, the types of attack and the normal class. The attack data then move forward to stage 4, receiving the flag of its specific attack class. If the data is found now to be from a normal class, the data is sent to stage 3. 


Stage 3 consists of verifying the data flagged in stages 1, 2, or 4, as normal data. In this stage, a Deep Learning model is applied to classify the data in a multiclass way, among the different classes of the data.
Although the input data from this stage has been previously flagged as normal data, a second step is needed to reduce the possibility of false positives and false negatives. Thus, the chosen methodology consists of applying an LSTM, which requires a robust model, capable of classifying the data more accurately. At the end of stage 3, we have: True Negatives ($TN'$), normal data correctly verified as belonging to the normal flow of the network; True Positives ($TP'$), data from attacks that were previously flagged as normal, but which stage 3 has verified are attacks; False Positives ($FP'$), normal behavior data that was incorrectly classified as an attack by stage 3; False Negatives ($FN'$), attacks that were not detected by stage 3.


Stage 4 receives the data considered to be from attacks in Stage 2, separated by attack class. Thus, this stage serves to verify the real belonging of the data to this class in order to reduce the number of false positives. At this stage, each attack has a specific model, which will verify if this data really belongs to that type of attack, or if it is a false positive. Thus, at this stage, \textit{Ensemble} classifiers are used to generate a specific and accurate model for each type of attack. The RF, kNN and SVM classifiers are applied to the stage 4 data and perform the classification based on the majority vote of their predictions. At the end of stage 4, the data is classified by type of attack, and thus we have: True Positives ($TP''$), attacks correctly classified in their attack class; False Positives ($FP''$), attacks incorrectly classified between attack classes, False Negatives ($FN''$), attacks incorrectly classified as normal class.

Thus, to obtain the results of the total system performance, the values of
Total True Positive ($TP_t$), Total False Positive ($FP_t$), Total True Negative ($TN_t$), Total False Negative ($FN_t$), are calculated following equations 1, 2, 3 and 4:

     \begin{equation} \label{vpt}
 TP_t	 = TP' + TP''
	\end{equation}
	 \begin{equation} \label{fpt}
FP_t	 = FP' + FP''
	\end{equation}
	\begin{equation} \label{vnt}
TN_t	 = TN'
	\end{equation}
	\begin{equation} \label{fnt}
FN_t	 = FN'+FN''
	\end{equation}

The $TP_t$, Equation 1, and the $FP_t$, Equation 2, are related to the stages where the attack class is assigned, and this classification can be correct ($TP_t$) or incorrect ($FP_t$). On the other hand, $TN_t$, Equation 3, and $FN_t$, Equation 4, are related to normal class assignment, both for instances truly belonging to the normal class ($TN_t$), and for attack data that were not correctly identified. ($FN_t$). Based on these equations the following metrics are calculated: Accuracy (ACC), Precision (Prec), True Positive Rate (TPR), False Alarm Rate (FAR ) and F1 score (F1).

It is important to note that in the proposed system, the data is sent in a continuous flow between the stages. As an instance is labeled, it is sent to the next part of the system, so it is not necessary to finish processing all the instances from one stage to start another. Thus, the system finishes when all instances of the dataset are labeled. Depending on the time required in each stage, this can be either at the end of stage 3 or at the end of stage 4. 

\subsection{The Network  point of view}

As Figure 1 shows, the system can be distributed in the network, each stage from the ML pipeline being part of a network layer.

\textbf{IoT layer}: This layer has a great diversity of IoT devices, such as sensors, smart lamps, and cameras, among other applications. These devices generate network traffic that serves as input to the system \textit{(a)}.

\textbf{Edge layer}: In this second layer we have the edge routers, which receive the network data flow coming from the IoT layer.
In edge routers, the first stage of the pipeline is implemented, the binary stream  ML attack detection. As This stage does not require much processing power, it can be implemented in SDN edge routers.
The data considered as normal traffic goes to the cloud layer \textit{(b)}, while data that was considered as attack data is sent to the fog layer \textit{(c)}.

\textbf{Fog Layer}:
This layer consists of more robust devices such as IDS and IPS that can implement stage 2 of the ML pipeline. After this layer, the data is sent to the cloud layer, flagged either as a normal \textit{(d)} or an attack data \textit{(e)}.

\textbf{Cloud Layer}:
At the cloud layer, remote data centers are then used to apply stage 3 and stage 4 of the ML pipeline system. With more processing and memory capacity it is possible to implement the Deep Learning and the Ensemble stages of the ML pipeline.

\section{Evaluation and Results}
\subsection{Experiment and Datasets}

To evaluate the proposed system, we use three network security datasets: BoT-IOT, TON-IOT, and CIC-IOT-2022. These datasets contain IoT network traffic data, both from normal behavior and from different types of attacks, and are among the most used in the context of detection and classification of IoT attacks in using ML techniques \cite{surveybases}. The following is a brief description of the three datasets:

 \textbf{BOT-IOT}: The dataset have 13,428,602 instances between data related to the normal network behavior and four types of attacks: \textit{Denial of Service Attack} (DoS), \textit{Distributed Denial of Service Attack} (DDoS), \textit{Information Theft Attack} (Theft), \textit{Reconnaissance} (Recon)  attacks.

   \textbf{TON-IOT}: It contains 3,202,452 instances related to normal behavior and 9 different types of attacks: \textit{Backdoor}, \textit{DoS}, \textit{DDoS}, \textit{Injection}, \textit{Man-in-the-middle} (MITM), \textit{Password}, \textit{Ransomware}, \textit{Scanning}, \textit{XSS}.

   \textbf{CIC-IOT-2022}: The dataset has 373,988 instances and aggregates data regarding the normal behavior of the IoT network and 4 classes of attacks:
  \textit{UDPflood}, \textit{TCPflood}, \textit{HTTTPflood} and \textit{Real Time Streaming Protocol Brute Force Attack} (RTSP).
  

All three datasets we pre-process from their original raw files (\textit{.pcap}), and then converted to a \textit{.csv} file using the CICFlowMeter toll \cite{cicflowmeter}, with aggregates network packets by flows with IP and Port from the same origin/destination. This tool is used in network benchmarks \cite{surveybases} \cite{cid17} to give the same feature set to all datasets used in the experiment. Thus all three datasets have contains the 81 flow features provided by \cite{cicflowmeter}. The datasets were used as input for the proposed system, read as an online and continuous flow of data. The experiments were carried out using a machine with an Intel Core i5-5200U processor with 2.20 GHz and 8 GB of available RAM, using the Windows 10 x64 operating system.

\subsection{Results}
This section presents the results obtained with the proposed system in the three databases presented in the case study. Table I presents the result obtained at the end of the data classification process in the BOT-IOT, TON-IOT, and CIC-IOT-2022 databases, according to equations 1, 2, 3, and 4.

In Table I, we can observe that the proposed system had a performance above 90\% in the evaluated metrics, in all the databases used in the case study. Also, the value of false alarms is less than 10\% in all cases. The system had similar Prec and TPR between BOT-IOT (96.76\%) and IOT-22 (96.97\%) datasets, however, the accuracy and F1 in the base BOT-IOT (98.91 \% and 98.46\%) is higher than the ones obtained in IOT-22 (96.55\% and 97.16\%). Thus, we have that the system had its best performance on the BOT-IOT dataset, followed by IOT-22, and finally on TON-IOT dataset, which had the lowest values in the evaluated metrics.

\begin{table}[ht]
   \begin{center}
    \caption{OMINIACS final results}
    \label{tab:geral}
\begin{tabular}{llllll}
\hline
\textbf{Dataset} & \textbf{ACC\%}	& \textbf{Prec\%}	&	\textbf{TPR\%}
&	\textbf{FAR\%} 	&	\textbf{F1\%}\\
\hline
BOT-IOT	&	98.91	&	96.76	&	99.88	&	03.24	&	98.46	\\
TON-IOT	&	94.33	&	90.45	&	99.83	&	9.55	&	94.35	\\
IOT-22	&	96.55	&	96.97	&	99.89	&	03.03	&	97.16	
\end{tabular}
\end{center}


   
   \begin{center}
    \caption{Results by stage: BOT-IOT Dataset}
    \label{r:nsl}
    
\begin{tabular}{clllll}
\hline
\textbf{Stage} & \textbf{ACC\%}	& \textbf{Prec\%}	&	\textbf{TPR\%}
&	\textbf{FAR\%} 	&	\textbf{F1\%} \\
\hline
\textbf{1}	&	72.77	&	99.93	&	72.63	&	00.07	&	84.12	\\
\textbf{2}	&	97.14	&	99.47	&	97.65	&	00.53	&	98.55	\\
\textbf{3}	&	98.02	&	98.01	&	99.96	&	01.99	&	98.98	\\
\textbf{4}	&	99.26	&	99.62	&	99.63	&	00.38	&	99.62	\\

\end{tabular}
\end{center}

   \begin{center}
    \caption{Results by stage: TON-IOT dataset}
    \label{tab:r:unsw}
\begin{tabular}{clllll}
\hline
\textbf{Stage} & \textbf{ACC\%}	& \textbf{Prec\%}	&	\textbf{TPR\%}
&	\textbf{FAR\%} 	&	\textbf{F1\%} \\
\hline

\textbf{1}	&	89.78	&	99.89	&	85.07	&	00.11	&	91.89	\\
\textbf{2}	&	84.96	&	98.06	&	86.38	&	01.94	&	91.85	\\
\textbf{3}	&	99.48	&	98.32	&	99.50	&	01.68	&	98.91	\\
\textbf{4}	&	99.75	&	99.81	&	99.91	&	00.19	&	99.86	\\
\end{tabular}
\end{center}


   \begin{center}
    \caption{Results by stage: CIC-IOT dataset}
    \label{r:c17}
\begin{tabular}{clllll}
\hline

\textbf{Stage} & \textbf{ACC\%}	& \textbf{Prec\%}	&	\textbf{TPR\%}
&	\textbf{FAR\%} 	&	\textbf{F1\%} \\
\hline

\textbf{1}	&	81.37	&	99.96	&	81.21	&	00.04	&	89.62 \\
\textbf{2}	&	97.84	&	97.85	&	99.55	&	02.15	&	98.91 \\
\textbf{3}	&	99.59	&	99.56	&	99.78	&	00.44	&	99.78 \\
\textbf{4}	&	99.37	&	99.90	&	99.45	&	00.10	&	99.68 \\

\end{tabular}
\end{center}
\end{table}

Tables II, III, and IV present the results for each stage of the system. In these tables, it is possible to observe the specific performance of each stage of the proposed system, in terms of the evaluated metrics. In general, we can observe that the ACC, Prec, TPR, and F1 results obtained in most stages are above 80\%, and that the FAR results are also below 3\%. An exception to this is stage 1 of the BOT-IOT database, Table 4, which had ACC (72.77\%) and TPR (72.63\%) less than 80\%, these being the worst results we obtained.

In addition, we can observe in tables II, III, and IV, the tendency of the metrics ACC, Prec, and TPR to have a lower value in stage 1, which increases in stages 2 and 3, and has a reduction in stage 4. This is due to the behavior of the ML methods used in each step of the system. In Stage 1, we have a binary classification system in a stream of data, which provides real-time data analysis. resulting in lower values in the evaluated metrics. Stage 2 is a multiclass stream classification, which no longer receives most of the data considered normal, thus having a better performance than the binary system of Stage 1. In stages 3 and 4, more accurate ML methods are used, deep learning LSTM and ensemble-based ML, giving better classification results.  

Tables V, VI, and VII present the results obtained by type of traffic class, for each dataset. These tables highlight the detection results of each class and the false alarms generated by the system for that specific class. In Table VI, we have the performance by class type for the BOT-IOT dataset. We can see that the normal class has the highest TPR (98.21\%), followed by \textit{DoS}, \textit{DDoS}, \textit{Recon} and \textit{Theft}. This follows the trend of classes with more data available having better results in terms of detection. On the other hand, the \textit{Theft} attack class had a significantly higher FAR (19.12\%) and lower TPR (82.30\%) than the others class, this may be a result of the confusion between this class with others such as DDoS and the Normal class that also have a large number of instances available.


\begin{table}[h]
   \begin{center}
    \caption{Results by Attack Class: BOT-IOT dataset}
    \label{tab:table1}
\begin{tabular}{llll}
\hline
\textbf{Class} &	\textbf{TPR\%} &	\textbf{FAR\%} & \textbf{Instances}	 \\
\hline

DoS	&	97.57	&	01.66	 & 89,246	  \\
DDoS	&	97.17	&	00.49	& 4,909,405	\\
Theft	&	82.30	&	19.12	& 4,913,920	\\
Reconn	&	95.90	&	00.05	 & 1,701	\\
Normal	&	98.21	&	4.86	 &3,514,330	\\

\end{tabular}
\end{center}


   \begin{center}
    \caption{Results by Attack Class: TON-IOT}
    \label{tab:table1}
\begin{tabular}{lllll}
\hline
\textbf{Class} &	\textbf{TPR\%} &	\textbf{FAR\%} & \textbf{Instances} \\
\hline

Backdoor	&	91.73	&	02.75 & 27,145	\\
DoS	&	92.41	&	02.90		 &145 \\
DDoS	&	95.54	&	03.98 &202 	\\
Injection	&	98.49	&	00.55	 &277,696	\\
MITM	&	97.29	&	0.54	 &517	\\
Password	&	97.00	&	15.9	&340,208	\\
Ransomware	&	68.65	&	37.02	 &5,098	\\
Scanning	&	82.31	&	7.01	 &36,205
\\
XSS	&	87.23	&	0.32	 &2,149,308	\\
Normal	&	98.82	&	1.18	 &2,515,236	\\

\end{tabular}
\end{center}


   \begin{center}
    \caption{Results by Attack Class: CIC-IOT-2022}
    \label{tab:table1}
\begin{tabular}{llll}
\hline
\textbf{Class} &	\textbf{TPR\%} &	\textbf{FAR\%} & \textbf{Instances}	\\
\hline

UDPflood	&	93.33	&	00.85 & 7,136	\\
TCPflood	&	89.99	&	00.95 & 28,560	\\
HTTPflood	&	98.14	&	00.28 &327,496	\\
RTSP	&	74.05	&	00.19 & 6,914	\\
Normal	&	99.64	&	07.20 & 3,882	\\

\end{tabular}
\end{center}

   \begin{center}
    \caption{Time taken to process data in OMINACS system: by each stage and by each dataset.}
    \label{tab:table1}
\begin{tabular}{llll}
\hline
\textbf{Stages}	&	\textbf{BOT-IOT}	&	\textbf{TON-IOT}	& \textbf{CIC-IOT-2022}  \\
\hline
	
\textbf{Total Time} &	2,984 \textit{(s)}	&	843 \textit{(s)}	&	91 \textit{(s)}	\\
\textbf{Stage 1} &	310 \textit{(s)}	&	155 \textit{(s)}	&	22 \textit{(s)}	\\
\textbf{Stage 2} &	790 \textit{(s)}	&	182 \textit{(s)}	&	39 \textit{(s)}	\\
\textbf{Stage 3} &	2,686 \textit{(s)}	&	758 \textit{(s)}	&	82 \textit{(s)}	\\
\textbf{Stage 4} &	2,201 \textit{(s)}	&	680 \textit{(s)}	&	76 \textit{(s)}	\\

\end{tabular}
\end{center}
\end{table}

Table VI shows the result by type of attack of the TON-IOT dataset. Again the Normal class, having the largest amount of data available, had one of the higher TPR rates (98.82\%). However, the XSS class, that have the second largest number of instances, had one of the lowest TPR results (87.23\%). In addition to this, the ransomware and the scanning attacks also had low TPR results (68.65\% and 82.31\%), with significantly high false alarm rates. These results might help explain the fact that the TON-IOT dataset had the lowest accuracy in this experiment, as shown in Table I.

Table VII shows the performance by traffic class for the CIC-IOT-2022 dataset. As in the other datasets, the Normal class also had the best TPR results (99.64\%), but in this case without having the largest amount of data. This might have caused the normal traffic class to have a higher FAR (07.20\%). The \textit{HTTPflood} class that had the more number of instances got the second higher TPR (98.14\%), followed by the \textit{UDPflood} (93.33\%), \textit{TCPflood} (89.99\%) and the \textit{RTSP} (74.05\%), all of them having a FAR lower than 1\%.

Therefore, considering all the results, we have that the normal traffic class had the best detection in all three datasets, with the highest  TPR and lowest FAR. This is mainly due to stage 3, where the LSTM is used to verify the data considered normal in stages 2, 3, and 4. This factor contributes to the reduction of false positives in the system since most of the normal traffic class is correctly classified.

Table VIII shows the time, measured in seconds, that the input data take to go through the system. For each dataset, it is shown the total time required to detect and classify network traffic data and the time spent in each stage of the system. We can see that the time required to process the BOT-IOT data is much higher than for the other datasets. This is due to the number of instances in this dataset BOT being much larger than the other datasets. Also, we can see that the CIC-IOT-2022 is processed very fast (in only 91 seconds) compared with the other datasets.

In addition, we can observe that Stage 3 has the determining time for the total system time, being the most time-consuming stage of the system. This is because at this stage the data classification process is carried out using the LSTM, a deep learning approach, which tends to be a more time-consuming method than the ones used in the other stages. However, as the results of tables II, III and IV demonstrate, this step has a high performance, with is reflected mainly in the high value of TPR and low FAR of the Normal class. Thus, although it is the most time-consuming stage, this stage is fundamental to the proposed system.

\section{Conclusion and Future Works}

Detecting online network attacks is still a major challenge in today's IoT networks. This paper tackles this issue by proposing an online and ML-Based system named OMINACS. With both an ML pipeline and IoT network view, OMINACS distributes four stages of attack detection and classification through the network.
We tested OMINACS with three IoT datasets achieving accuracy and precision results above 90\%, with a reduced false alarm rate. We also were able to analyze the results by each stage and by each attack class. Therefore, the results confirm that the system is able to detect and classify a large variety of network attacks. 
In future work, the authors aim to test the proposed system in a production network, such as in a small campus network where the scalability of the system can be evaluated, allowing further improvements on both the ML pipeline and the IoT network view.


\section*{ACKNOWLEDGMENT}
The authors would like to thank the São Paulo Research Foundation (FAPESP) for the financial support through Grant 2020/04031-1.

\bibliographystyle{IEEEtran}
\bibliography{references}

\end{document}